\documentclass[%
superscriptaddress,
 amsmath,amssymb,
 aps,
 reprint,
prb,
floatfix,
]{revtex4-2}
\usepackage{graphicx,xcolor}
\usepackage{multirow}
\definecolor{darkblue}{RGB}{0,0,150}
\definecolor{nightblue}{RGB}{0,0,100}

\usepackage{mathrsfs,dsfont,mathtools}

\usepackage{ulem}

\usepackage{dcolumn}% Align table columns on decimal point
\usepackage{bm}% bold math
\usepackage[
colorlinks,
citecolor=darkblue,
linkcolor=darkblue,
urlcolor=nightblue]{hyperref}% add hypertext capabilities
\usepackage[english]{babel}
\usepackage[babel,kerning=true,spacing=true]{microtype}
\usepackage{pifont}
\usepackage{feynmp-auto}

\newcommand{\bbronze}{K\textsubscript{0.3}MnO\textsubscript{3}}
\AtBeginDocument{\renewcommand{\natexlab}[1]{#1}}% <--- the fix
\begin{document}

\title{
Optically-Induced Faraday-Goldstone Waves 
}
\author{Daniel Kaplan}
\email{d.kaplan1@rutgers.edu}
\affiliation{Center for Materials Theory, Department of Physics and Astronomy, Rutgers University, Piscataway, New Jersey 08854, USA}
\author{Pavel A. Volkov}
\affiliation{Department of Physics, University of Connecticut, Storrs, Connecticut 06269, USA}
\author{Andrea Cavalleri}
\affiliation{Max Planck Institute for the Structure and Dynamics of Matter, 22761 Hamburg, Germany}
\affiliation{Department of Physics, Clarendon Laboratory, University of Oxford, Oxford OX1 3PU, United Kingdom}
\author{Premala Chandra}
\affiliation{Center for Materials Theory, Department of Physics and Astronomy, Rutgers University, Piscataway, New Jersey 08854, USA}
\affiliation{Center for Computational Quantum Physics, The Flatiron Institute,162 5th Avenue, New York, NY 10010}
\date{\today}
\begin{abstract}
\noindent Faraday waves, typically observed in driven fluids, result from the confluence of nonlinearity and parametric amplification.
Here we show that optical pulses can generate analogous phenomena that persist much longer
than the pump time-scales in ordered quantum solids.  
We present a theory of ultrafast light-matter interactions within 
a symmetry-broken state; dynamical nonlinear coupling between the Higgs (amplitude) and the Goldstone (phase) modes drives an emergent phason texture that oscillates in space
and in time: Faraday-Goldstone waves. Calculated signatures of this spatiotemporal order compare well with measurements
on \bbronze; Higgs-Goldstone beating, associated
with coherent energy exchange between these two modes, is also
predicted.
We show this light-generated crystalline state is robust to thermal noise, even when the original Goldstone mode is not. Our results offer a new pathway for the design of periodic structures in quantum materials with ultrafast light pulses.
\end{abstract}

\maketitle

\section{Introduction}
{\allowdisplaybreaks

Pattern formation in driven systems has been studied
extensively, particularly in classical and quantum fluids \cite{Faraday1831,Cross93,Nguyen19,Liebster25}.
An early example, discussed by Faraday \cite{Faraday1831}, occurs when a liquid is vertically vibrated. Above
a threshold frequency, its surface displays a dynamical instability to a standing wave pattern that oscillates at half the drive frequency; these
Faraday waves result from the parametric excitation of surface collective modes \cite{Faraday1831,Cross93,Nguyen19,Liebster25}.
In a modern context, in solids, one may ask whether excitations similar to those proposed by Faraday can be made possible, e.g., with light; it is known that ultrafast optics offer great promise towards realizing new dynamical phases
\cite{disa2021engineering,bao2022light,Giannetti_2016,de2021colloquium}.

In pump-probe experiments, pulsed excitations
of systems close to criticality often drive phase transitions;
with strong fluence they can also melt underlying ordered
ground-states
\cite{ginzburg1963scattering,Liu13,schafer2010disentanglement,schaefer2014collective,tomeljak2009dynamics,kung2013time,fu2020nanoscale,dolgirev2020amplitude,zong2019evidence,Rausch2019Auger,huber2014coherent,Zong2019slowingdown,kogar2020light,ilyas2024terahertz,yang2025photoinduced, Masoumi2025mestability,forst2011nonlinear,mitrano2016possible,zhuang2023light,yusupov2010coherent,Buzzi21}. An outstanding question is whether ultrafast optical pulses can generate new out-of-equilibrium orderings, given recent work on the generation of such order from continuous wave excitations 
\cite{kiselev2024inducing,Kaplan2025,kaplan2025spatiotemporalorderparametricinstabilities,kiselev2025,hosseinabadi2025}.
More specifically we explore whether, in analogy with Faraday
waves, pulsed excitations can induce parametric resonance
that leads to spatiotemporal patterns.
We address this issue using a combination of computational and analytic approaches, producing signatures that are
consistent with experimental results. 
We end with specific predictions for
future measurements and then discuss broader implications
of our work.

We consider a system with a continuous broken symmetry characterized by a gapped (Higgs) amplitude and
a gapless (Goldstone) phase, oscillations in
the magnitude and direction of the order parameter respectively;
in a minimalist model, these two modes form the system's
dynamical degrees of freedom \cite{pekker2015amplitude}. Light pulses have been shown to induce coherent, ${\bf q}=0$ oscillations of the Higgs mode ($\mathbf{q}$ being the momentum of the excitation); by contrast, the possibility of inducing coherent ${\bf q}\neq0$ ordering with ${\bf q}=0$ light is not straightforward, given that any coupling to the phase mode must be indirect \cite{PWA63,Juraschek20}. 
We present a mechanism for generating incommensurate
dynamical order by a $\mathbf{q}=0$ 
a pulsed excitation of the Higgs mode; it then {\sl coherently} excites 
phasons at half the Higgs mode frequency and finite wavevector.
Above a threshold fluence, pulsed excitations deep in the 
ordered phase generate symmetry-breaking modulations in space and time through a mechanism defined by a dynamical nonlinear Higgs-Goldstone coupling. 

There are several experimental consequences of our theory.
We show that the amplitude mode spectral intensity saturates above the threshold for parametric Goldstone mode generation ;this quantity is directly observable by reflectivity and
other $\mathbf{q} = 0$ light probes.
Because of the inherent nonlinearities of the theory, the bare frequency of the amplitude mode (at $\mathbf{q}=0$) is dynamically renormalized and softens. In the low dissipation regime, we predict coherent Higgs-Goldstone beating patterns involving the coherent exchange of energy
between phase and amplitude modes that should be optically accessible.
Exploring the temperature-dependence of the pulse-induced
spatiotemporal ordering, we find it is only weakly affected by the presence of thermal noise,
even though 
the $\mathbf{q}=0$ gapless Goldstone mode exhibits significant thermal drift.

Our theoretical results compare well with ultrafast pump-probe measurements on \bbronze  (blue bronze), a prototypical 1D charge density wave (CDW) host \cite{pouget1985structural,schutte1993incommensurately,schlenker2012low,Gruner_CDW}. More specifically we show excellent fits to reflectivity data at $T=20K$ deep in the CDW phase \cite{Mankowsky2017};
furthermore time-resolved ARPES measurements point to the existence of half-frequency oscillations of momenta away from the $2k_F$ instability generating the CDW in \bbronze \cite{Liu13}. 

Our approach is widely applicable to pump-probe experiments; it generalizes earlier proposals for out-of-equilibrium spatiotemporal order \cite{Kaplan2025,kaplan2025spatiotemporalorderparametricinstabilities} to the experimentally relevant  ultrafast pump-probe setup \cite{disa2021engineering}. Furthermore our results point to the coexistence of coherent non-thermal and incoherent thermal states, implying new possibilities in the design of light-induced exotic phases in quantum materials.}

\section{Symmetry-Breaking Order Out of Equilibrium}
\label{sec:theory_intro}
Here we study the dynamics of a complex order parameter $\Psi$ of a system with spontaneously broken continuous $U(1)$ symmetry \cite{hohenberg2015introduction}. 
Following previous dynamical studies of density wave systems \cite{schaefer2014collective,dolgirev2020amplitude}),
we assume a relaxational approach to equilibrium  \cite{hohenberg-halperin}:
\begin{equation}
    \frac{\delta \mathcal{F} [\Psi,\Psi^*]}{\delta \Psi} = -\eta \Psi + \mathcal{R}(x,t),
    \label{eq:dyn}
\end{equation}
where $\eta$ is the system's relaxation rate and $\mathcal{R}(x,t)$ is the thermal noise satisfying the fluctuation-dissipation theorem $\langle \mathcal{R}(x,t) \mathcal{R}(x',t') \rangle = \eta T \delta(x-x')(t-t')$. For the system's free energy, $\mathcal{F} [\Psi,\Psi^*]$, we include both the static energy contribution, corresponding to the phenomenological Landau theory,  and a dynamical term, describing the kinetic energy of the order parameter, i.e. 
\begin{align}
    \mathcal{F} = \mathcal{F}_\textrm{static}+\mathcal{F}_\textrm{dynamical},
    \label{eq:ftot}
\end{align}
where
\begin{align}
    \mathcal{F}_\textrm{static} = \int_{x,t}\left\lbrace\frac{v^2}{2}|\partial_x\Psi|^2+\frac{\alpha}{2}|\Psi|^2+\frac{\beta}{4}|\Psi|^4\right\rbrace.
    \label{eq:fstat}
\end{align}
and
\begin{align}
    \mathcal{F}_\textrm{dynamical} =-  \int_{x,t} \partial_t\Psi^{*}\partial_t\Psi.
    \label{eq:fdyn}
\end{align}
The presence of $\mathcal{F}_\textrm{dynamical}$ is known
in charge density wave (CDW)  systems \cite{dolgirev2020amplitude,yusupov2010coherent} due to inertia of ionic motion; it also arises in superconductors at low temperatures \cite{abrahams1966time,pekker2015amplitude}
though a there full description of the order parameter dynamics must involve quasiparticles above the gap \cite{stoof1993,shimanoarcmp} in the absence of additional order parameters \cite{lit_varma}. We thus focus our
attention on charge- and spin-density wave systems where 
Eqs.~\eqref{eq:dyn}-\eqref{eq:fdyn} provide a good phenomenological description of the dynamics. 

The order parameter can be represented as $\Psi(x,t) = \Delta(x,t)e^{i\phi(x,t)}$, where $\Delta$ and $\phi$ are its (real) amplitude and phase respectively. The equations of motion for $\Delta$ and $\phi$ are then obtained from the variational principle $\frac{\delta \mathcal{F}}{\delta \Psi} =0,\frac{\delta \mathcal{F}}{\delta \Psi^{*}} =0 $ (see Appendix \ref{App:A0}) that yields two coupled equations,
\begin{align}
    -v^2 \partial_x (\Delta^2\partial_x\phi)+\partial_t(\Delta^2\partial_t\phi) + \eta \Delta^2\partial_t\phi = \Delta \mathcal{R}_\phi
    \label{eq:phase_eq}
\end{align}
\begin{align}
    \notag \left[-v^2 \partial_x^2+\partial_t^2+\eta\partial_t \right]\Delta &+ \\  \Delta(
    %2 |\alpha|
    \alpha + D(\phi)) &+ \beta \Delta^3 = \mathcal{R}_\Delta.
    \label{eq:amp_eq}
\end{align}
where Eqs.~\eqref{eq:phase_eq} and ~\eqref{eq:amp_eq} 
describe the phase (Goldstone) and the amplitude (Higgs) 
modes respectively.
Here $D(\phi) \equiv v^2(\partial_x \phi)^2-(\partial_t\phi)^2$ is the dynamical potential of the phase acting on the amplitude mode; $\mathcal{R}_{\Delta,\phi}(x,t)$ is a noise source acting as a thermal bath that satisfies $\langle R_\alpha(x,t)R_\beta(x',t')\rangle=\eta T\delta_{\alpha\beta}\delta(x-x')\delta(t-t')$, where $T$ is the temperature and $\alpha,\beta = \{\Delta,\phi\}$ (see Appendix \ref{App:A0} for the derivation).
Linearizing the equations in the vicinity of the equilibrium $\Delta \approx \Delta_0 = \sqrt{-\alpha/\beta}$, with $\Delta \to \Delta+\delta\Delta$ and $\phi \to \phi+\delta \phi$, we obtain (for $\eta=0$) the standard dispersions 
$\omega = \sqrt{2\alpha+v^2q^2}$ and $\omega=vq$ for the Higgs and
the Goldstone modes respectively.

We next study the impact of the nonlinear Higgs-Goldstone couplings in Eqs.~\eqref{eq:phase_eq} and \eqref{eq:amp_eq}
out of equilibrium; we note that they vanish when $\Delta$ and $\phi$ are time- and space-independent.
Motivated by experiment, we consider the pump's action on $\Psi$ to be instantaneous and therefore invoke the following initial conditions,
\begin{align}
    \Delta(x,0)/\Delta(0) = 1-\delta, ~~ \phi=0,
    \label{eq:init}
\end{align}
where we use the equilibrium value of $\Delta(0)=\sqrt{-\alpha/\beta}$  in the ordered phase $\alpha<0$.
%$A_1$
Though the amplitude mode is typically not optically active, belonging to a fully symmetric irreducible representation, 
pump-probe experiments can access it indirectly 
\cite{Zeiger92} by transferring energy from optically active modes to the Higgs mode. For such an excitation mechanism, we may expect $\delta$ to be proportional to the total pulse energy, i.e. fluence.

\begin{figure}[ht!]
    \centering
    \includegraphics[width=1.0\columnwidth]{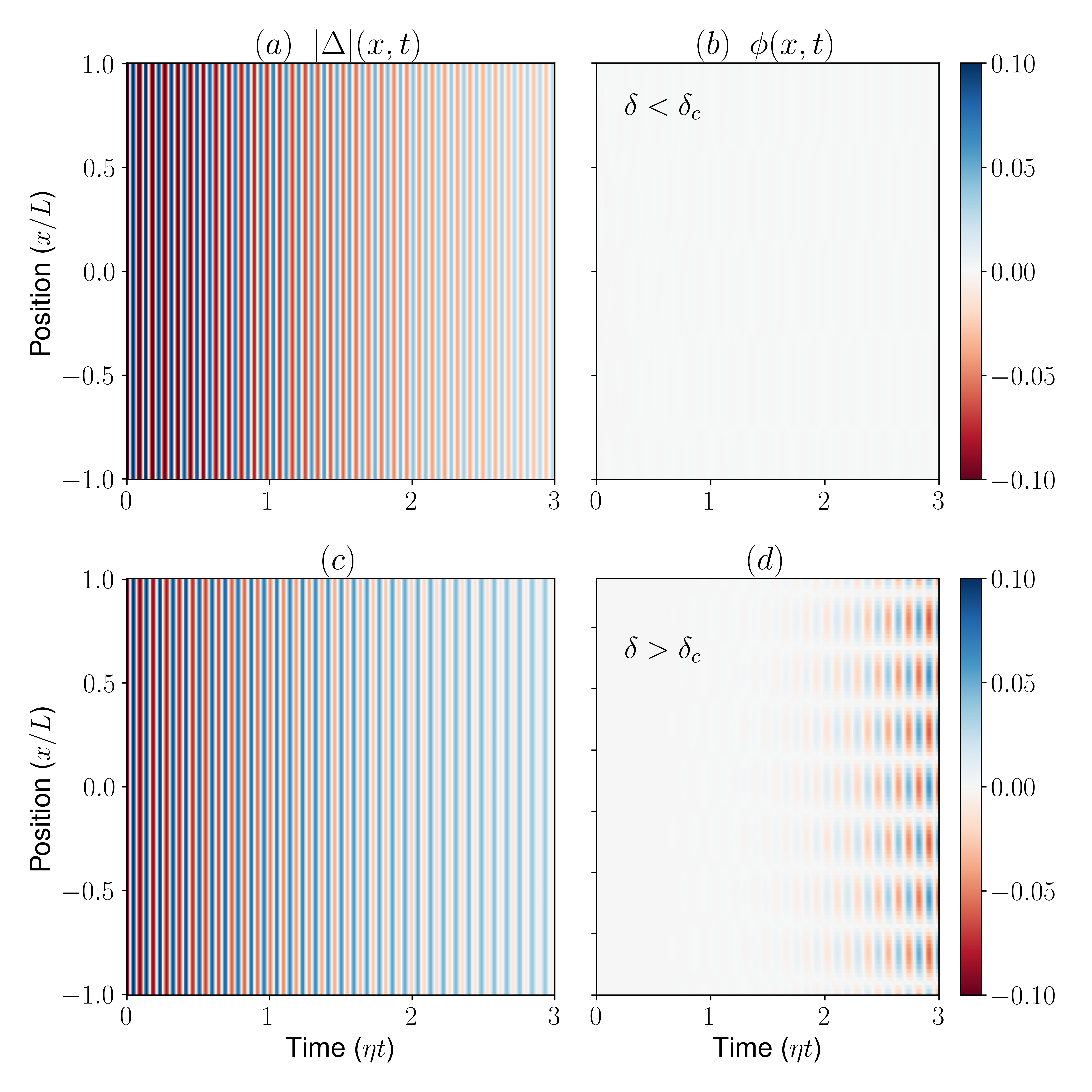}
    \caption{Spatial and temporal dependence of the order parameter's amplitude and phase after a pulse excitation, Eq. \eqref{eq:init}.
    (a,c) Spatial and temporal profile of $\Delta(x,t)$, below (a) and above (c) the the threshold $\delta$ for parametric excitations. Note that in below the threshold in (b), the phase mode does not respond at all and remains fixed at $\phi =0$. Above the threshold, in (d), the phase mode exhibits coherent spatial and temporal oscillations. Here, $\eta =0.5 m$.}
    \label{fig1}
\end{figure}

In order to gain intuition about the role of the Higgs-Goldstone couplings, we first solve Eqs.~\eqref{eq:phase_eq} and \eqref{eq:amp_eq} numerically (see App. \ref{app:num} for details).
In Fig.~\ref{fig1} we present results for $\Delta(x,t)$ and $\phi(x,t)$ for different initial displacements of the amplitude mode $\delta$
with $T = 0^+$. Here we find two contrasting regimes: in Figs.~\ref{fig1}(a-b), the driving (a) is sufficiently weak, and the phase mode does not respond (b), as its amplitude remains flat in both space and time; however 
above a threshold drive Figs.~\ref{fig1}(c-d) show the phase (in (d)) developing periodic oscillations both in time and space.

\section{Parametric Instability of the Goldstone Modes}
\label{sec:param}
In order to explain the spatial and temporal ordering we observe in numerical simulations (Fig.~\ref{fig1}), we analyze 
Eqs.~\eqref{eq:phase_eq} and \eqref{eq:amp_eq} with initial conditions Eq.~\eqref{eq:init} at $T=0$, assuming weak 
nonlinearity. Using the notation $\Delta = \Delta(0)+\Delta(x,t), \phi = \phi(x,t)$ and keeping the lowest order terms in $\Delta$, we obtain:
\begin{align}
    \notag & \Delta(0)^2 \left(-v^2 \partial_x^2 +\partial_t^2+\eta\partial_t\right)\phi + \\ & ~~~~ 2 \Delta(0)(-v^2 \partial_x \Delta \partial_x\phi+\partial_t\Delta \partial_t\phi) =0
    \label{eq:varphi}
\end{align}
and 
\begin{align}
    \left[-v^2 \partial_x^2+\partial_t^2 +\eta\partial_t\right] \Delta +m^2 \Delta = -|\Delta(0)|D(\phi).
    \label{eq:amp}
\end{align}
where $m^2=-2\alpha$.
We observe immediately that the phase and amplitude modes are coupled via gradient terms; furthermore they have a cubic coupling that follows from a potential  $V(\phi,\Delta) = |\Delta(0)|\int_{x,t}\Delta(x,t)[v^2(\partial_x \phi)^2-(\partial_t \phi)^2]$. The static and dynamical implications of such
a finite-momentum, finite-frequency cubic coupling
have been previously discussed for continuous harmonic
drives \cite{Kaplan2025,kaplan2025spatiotemporalorderparametricinstabilities}. 
Here we are studying a system that is {\sl not} continuously driven, but is rather excited by an instantaneous pulse represented by the initial condition \eqref{eq:init}.
Solving Eq.~\eqref{eq:amp} ignoring nonlinearities and the coupling $D(\phi)$, we obtain
\begin{align}
    \Delta(t)/|\Delta(0)| = 1-\delta e^{-\eta t/2}\cos(mt),
\end{align}
where we explicitly assumed $m \gg \eta$ and $|\delta| \ll 1$. Introducing this result to Eq.~\eqref{eq:varphi} and applying the spatial Fourier transform on $\phi$, we obtain
\begin{align}
    \ddot{\phi}_q+\eta\dot{\phi}_q +v^2q^2\phi_q -m\delta e^{-\eta t/2}\sin(mt)\dot{\phi_q} = 0,
    \label{eq:second_phi_q}
\end{align}
where we additionally neglected all terms higher order in $\eta$, using $m \gg \eta$ (see Appendix \ref{App:A0}). Eq.~\eqref{eq:second_phi_q} corresponds to a parametrically driven oscillator with a decaying drive and can be solved for $\delta, \eta\ll m$ using standard techniques \cite{landau2013mechanics}. Specifically, the solution is sought in the form ${\rm Re} \ \phi_q(t) = a_q(t)e^{i m t/2}+\textrm{c.c.}$ where $a_q(t)$ is a slow function $\dot a_q(t)/a_q(t)\ll m$. One can then obtain closed equations for $a_q(t)$ (see Appendix \ref{app:b} for details). For $ q = q_0 \equiv \frac{m}{2 v}$ one obtains:
\begin{equation}
\dot a_q  + \frac{\eta}{2} a_q - \frac{\delta}{4} e^{-\lambda t} a_q^*=0,
\label{eq:aq}
\end{equation}
and a complex conjugate equation for $a_q^*(t)$. The
\begin{equation}
a_q(t) = \frac{1+i}{2} a_+(0)  e^{\lambda_+(t)}
+
\frac{1-i}{2} a_-(0) e^{\lambda_-(t)},
\label{eq:sol_par_an}
\end{equation}
where $a_\pm(0) = [{\rm Re}a_q(0)+{\rm Im}a_q(0)]$ and:
\begin{equation}
        \lambda_\pm(t) =-\frac{\eta t}{2} \pm \frac{m \delta}{2}\frac{1-e^{-\eta t/2}}{\eta}.
        \label{eq:grate}
\end{equation}
While both components of \eqref{eq:sol_par_an} ultimately decay to zero at $t\to \infty$, the first one will exhibit initial growth as a function of time for $\delta > \frac{2 \eta}{m}$. This threshold matches the one for continuous driving \cite{Kaplan2025}. This growth persists only for a restricted period of time, until the drive decays below the dissipation.

As a measure of such parametric process strength, we consider the gain in the maximal value of $\phi_q(t) \propto \max |a_q(t)|$. The maximum gain is obtained at time $t_\textrm{max} = \frac{2}{\eta}\ln\frac{m \delta}{2\eta}$, and the resulting gain $G = \ln \frac{\tilde{a}}{\tilde{a}(0)} = \frac{m\delta}{2\eta}-1-\ln\frac{m \delta}{2\eta}$. A practical threshold is set by the requirement that $a_q(t_{max}) \sim a(0) e^G$ becomes larger than the noise floor or instrumental sensitivity, $a_{exp}$. This criterion will is expected to depend weakly on initial conditions due to the influence of thermal noise, i. e. $\delta_{exp} \approx \frac{2\eta}{m} \log[a_{exp}/a(0)]$ for $m \delta \gg \eta$; for low damping $\eta\ll m$ one expects that generically 
$\delta_{exp} \lesssim 1$, suggesting that the parametric Goldstone amplification should be observable in experiments.

\paragraph*{Amplitude Response and Frequency Softening---}
\label{sec:soft}
The theory for $\Delta$ is nonlinear even in absence of coupling to phasons.
 We note that the leading correction to Eq.~\eqref{eq:amp} is at order $\Delta^2$, which has the form $-3\sqrt{|\alpha|\beta}\Delta^2 = -3\beta |\Delta(0)|\Delta^2$. This corresponds to the standard form of a damped Duffing oscillator \cite{kovacic2011duffing,jordan2007nonlinear}
where it is known that the fundamental frequency ($\omega_0$) is approximately,
\begin{align}
    \omega_0 \approx m -\frac{3\beta|\Delta(0)| \delta_0^2}{2},
    \label{eq:soft}
\end{align}
where $\delta_0$ is the initial displacement. Intuitively, as the amplitude is displaced to where asymmetries in the Higgs potential increase its width, its classical turning time is also increased and its oscillation frequency decreases accordingly. This result is notably \textit{independent} of the parametric excitation and results from  the inherent nonlinearity of the Higgs potential in the Landau theory. 

\section{Implications}
We now study the full dynamics of the phase and amplitude modes, as expressed in Eqs.~\eqref{eq:phase_eq} and \eqref{eq:amp_eq}, at $T=0$. Here, we scale all relevant quantities with the bare parameters of the Landau theory. From the outset, the coefficient $\alpha$ is fixed, and this determines the bare mass $m^2=-2\alpha$. All dynamics are generated by modifying the initial condition $\Delta(x,0) = \delta_0/|\Delta(0)|$, that is, we measure the displacement from equilibrium in units of the equilibrium amplitude of the order. We then solve Eqs.~\eqref{eq:phase_eq} and \eqref{eq:amp_eq}, subject to the initial conditions $\partial_t\Delta(x,0)=0,\phi(x,0) = 0,\ \partial_t \phi(x,0) = \mathcal{U}$, where $\mathcal{U}$ is a uniform, randomly distributed collection of small initial values the size of our discretization in space; we then employ and average over $N=30$ realizations of this random initialization. The effect of finite dissipation $\eta$ is also examined in the following sections.

\subsection{Weak Pump - Linear Response}
When the displacement out of equilibrium is small, we expect to see only a linear response of the amplitude mode.
In Fig.~\ref{fig1}(a)-(b) we plot the spatial and temporal dependence of the amplitude and phase modes respectively. While oscillations are found in the amplitude, the phase remains fixed at $\phi=0$, which is its initial condition. This corresponds to the normal regime, where the Higgs and Goldstone modes are decoupled.
Focusing on particular cut in space, we plot the temporal oscillations in Fig.~\ref{fig2}(a). The phase mode, similarly, in Fig.~\ref{fig2}(b) is zero and independent of time for $\eta t \gg1$.
The Fourier transform of the oscillations reveals a sharp peak at the expected frequency of $\omega_0^2=m^2 = -2\alpha$ which is shown in Fig.~\ref{fig2}(c), in agreement with the general principles of linear response. In Fig.~\ref{fig2} (b,inset) we also plot the dependence of the $\Delta(q=0,\omega=m)$ amplitude on the initial displacement $\delta_0$. Here we observe linear response, as expected for small oscillations.

\subsection{Parametric Instability and Faraday-Goldstone Waves}
\label{sec:onset}
Beyond the linear regime, the coupling between the phase and amplitude modes becomes relevant.
\begin{figure}[ht!]
    \centering
    \includegraphics[width=1.0\columnwidth]{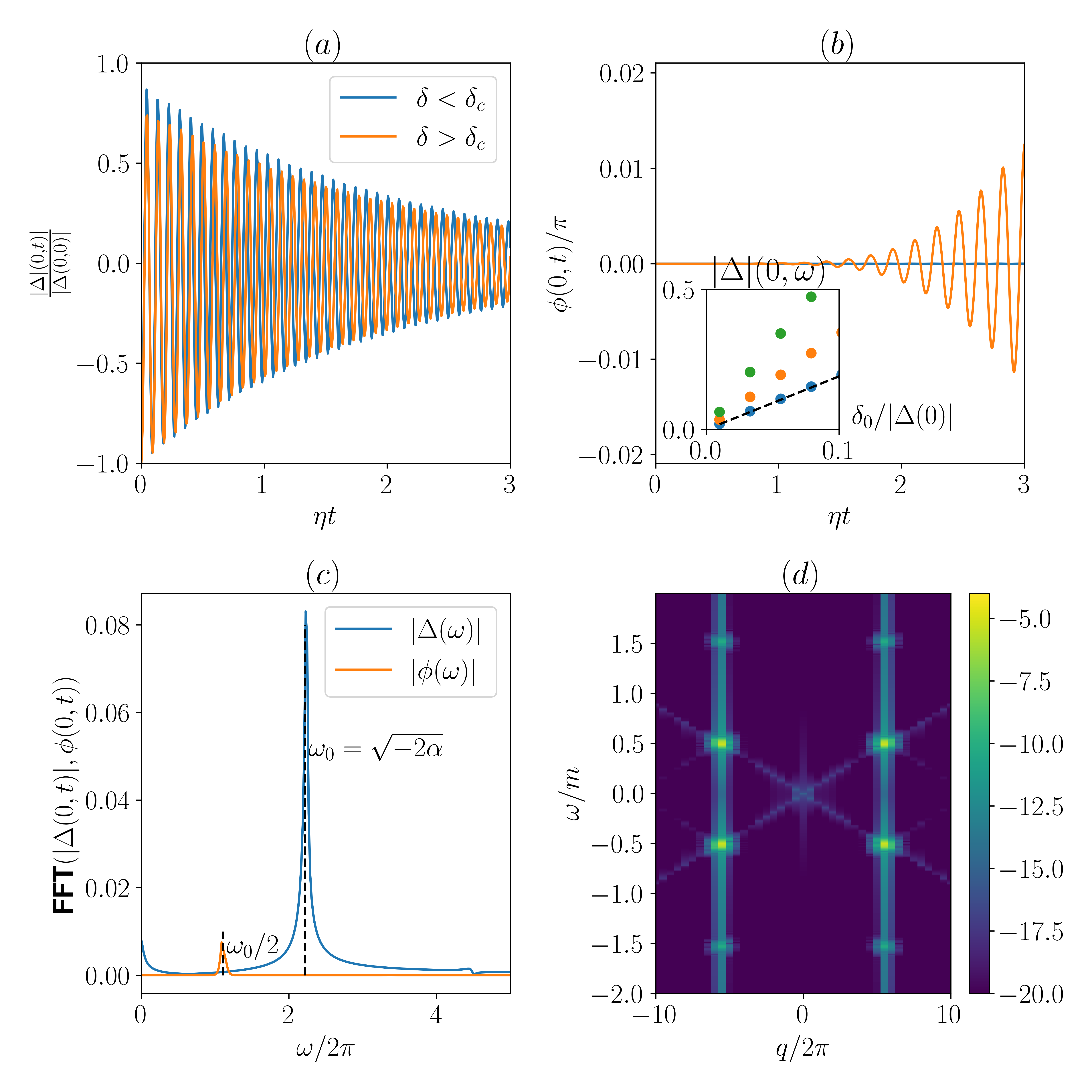}
    \caption{Onset of parametric oscillations in the out-of-equilibrium state. (a) shows temporal oscillations at the $x=0$ cut, for two regimes, above and below the threshold $\delta$ discussed in Sec.~\ref{sec:param}. (b) As in (a), but for the phase $\phi(x,t)$. While no significant change is seen for the amplitude mode $\Delta(x,t)$ (a), above a threshold a parametric instability sets in, manifest in a transient exponential growth in time. Inset shows the dependence of the spectral weight of the amplitude mode at $\omega=m$ on $\delta$ remaining linear for several dissipation values $\eta$ (green, orange, blue curves). (c) Fourier spectrum of the amplitude and phase modes (blue, orange, respectively).  Modes shows a clear resonance at the  $\omega=m$ for the amplitude, and the parametrically driven mode $m/2$ for the phase. (d) Fourier spectrum $\phi(q,\omega)$ for the phase mode. Clearly visible are the linear dispersion of the phase mode, the resonances at $\pm m/2, q=\pm q_0\approx \pm m/(2v)$ and higher-order replicas driven by nonlinearities.}
    \label{fig2}
\end{figure}
As shown in Sec.~\ref{sec:param} an instability to pairs of modes at frequency $\omega \sim m/2$ may occur above a threshold that is primarily determined by the coupling strength ($|\Delta(0)|$ ) and the dissipation ($\eta$). For large displacements, the phase mode develops coherent oscillations in both space \textit{and} time, as shown in Fig.~\ref{fig1}(b). The %precise dispersion relation
The Fourier spectrum of the phase mode in this state in frequency and momentum space is plotted in Fig.~\ref{fig2}(d). The latter shows the familiar linear dispersion expected of a Goldstone mode and sharp resonances near $m/2$, signifying the onset of order. In the 1D example considered here, modes appear in pairs of $\pm q_0, \pm m_0/2$. Additionally, for a cut in real space, the temporal Fourier transform (Fig.~\ref{fig2}(c)) again confirms the existence of a sharp peak at $m/2$.

\begin{figure}[ht!]
    \centering
    \includegraphics[width=1.0\columnwidth]{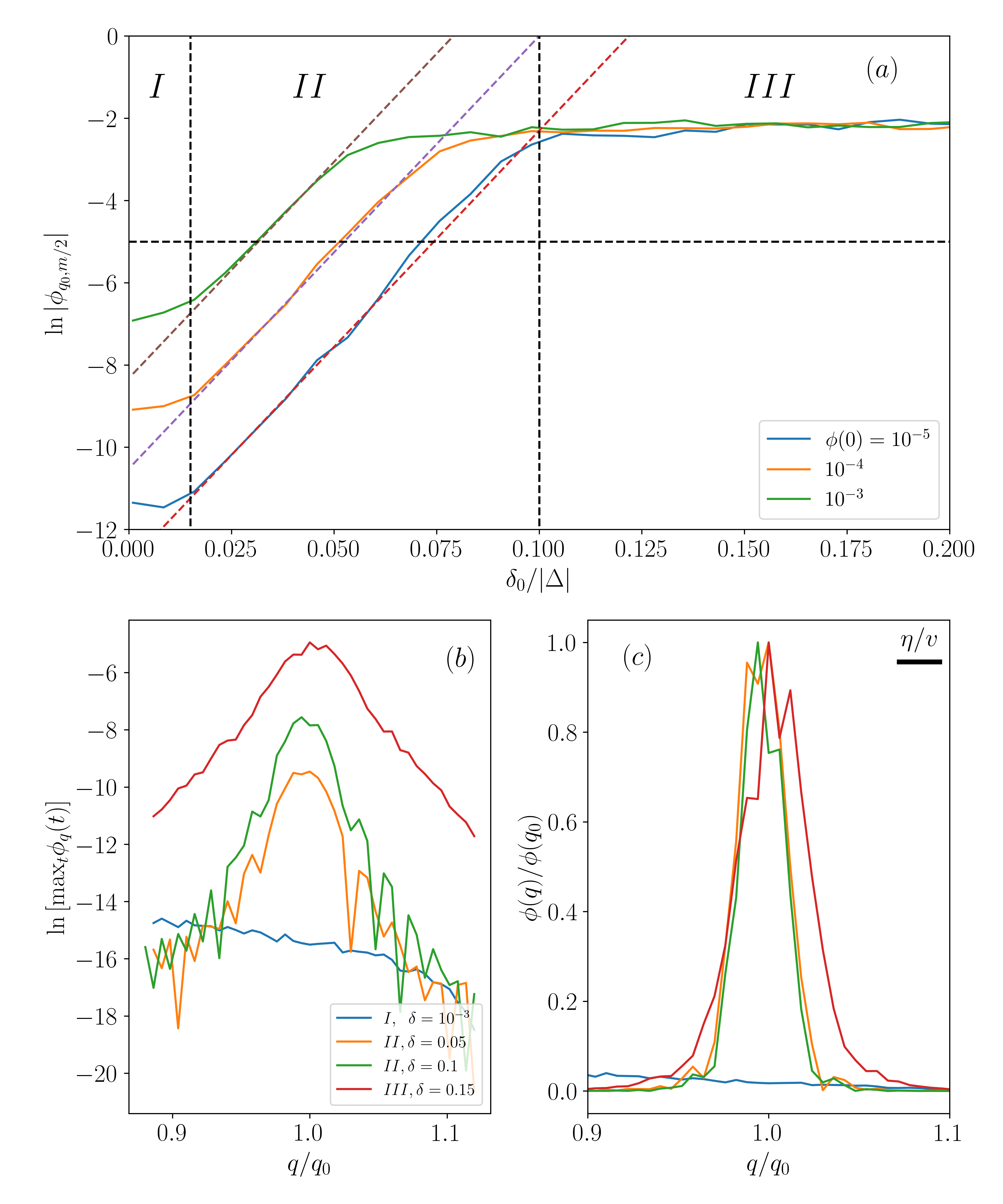}
    \caption{Threshold properties and momentum space structure of the Goldstone-Faraday waves. (a) Log of the Fourier amplitude $\phi(q_0,m/2)$ as a function of drive intensity for three different initial conditions $\phi(0) = 10^{-5}, 10^{-4},10^{-3}$. Colored dashed lines correspond to the growth rate $\lambda$ (Eq.~\ref{eq:grate}), off set only by the initial value of $\phi$. The different regimes $I,II,III$ (main text) are delineated. (b) Peak structure near $q_0$ at maximum gain $t_{\max}$, for different initial displacements $\delta$ on a log scale. In region $I$, no peak is visible, since it is below the fundamental threshold. The peaks intensity clearly grows with displacement $\delta$, showing that parametric amplification occurs for a range of $q$ modes. In region $III$ nonlinearities cause deviation from Gaussianity for the peak structure (red curve) 
    (c) Same as (b), but normalized on a linear scale. Even though a range of $q$ modes are parametrically excited, the overall similarity of peaks in region $II$ suggests that spatial coherence is maintained up to a minimal width of $\eta/v$ (black scale bar). In region $III$,  broadening increases. }
    \label{fig8}
\end{figure}

We now investigate the evolution of the Faraday-Goldstone instability in more detail.
In contrast to the continuous wave regime, the effect of a decaying impulsive excitation leads to different thresholds, which are more sensitive to the properties of the system. Firstly, there exists an \textit{absolute} threshold, set by the condition for the exponential growth of the parametrically driven Faraday wave, as presented in Eq.~\eqref{eq:grate}. This condition is independent of the initial state of the system, and is set by the mass of the Higgs mode $m$ and the dissipation $\eta$, leading to $\delta > \frac{2\eta}{m}$. Below this value no growth in time is expected. We plot this in Fig.~\ref{fig8}(a), where this region is denoted $I$. After crossing this threshold, the system enters into the exponential growth regime $II$, as expected from Eq.~\eqref{eq:grate}. In Fig.~\ref{fig8}(a), this is clearly visible on a log plot; however, as we explained in Sec.~\ref{sec:param}, the total parametric gain is sensitive to the initial state. Fixing the desired threshold by an empirical constraint, such as $\ln \phi_{q_0,m/2} =\ln 10^{-3}$, means that different initial conditions (the different curves in Fig.~\ref{fig8}(a)) reach the constraint at different displacements, depending on how large the initial value of $\phi_q$ was. However, the growth {\it rate} is universal, as we show by placing lines that correspond to $\delta-\frac{2\eta}{m}$, which is determined in Eq.~\eqref{eq:grate}. When the system is driven harder, the gain is cut off by the inherent nonlinearity of the theory. This is region $III$, where the intensity of the parametrically amplified mode saturates. 

It is instructive to study the properties of each of the regimes $I,II,III$ by considering the momentum space structure of the parametrically induced wave. in Fig.~\ref{fig8}(b), we show the fine structure of the resonance near $q_0$, evaluated at the time of maximum gain, $t_{max}$ (see Sec.~\ref{sec:param}). In regime $I$, as stated, no peak is observed at all. In regime $II$, we observe the clear parametric amplification for a range of modes in the vicinity of $q_0$.
As shown in Fig.~\ref{fig8}(b), the width of the peak in $II$ appears to be intact, while the peak in region $III$ is larger, and shows notable change in the shape. 
Remarkably, studying the peak structure on a linear scale (Fig.~\ref{fig8}(c), all curves normalized by their maximal value), we find that the width of the distribution does not change much, despite the exponential growth in regime $II$, but grows somewhat on entering the regime $III$.

These observations are in stark contrast to the case of continuous driving, where resolution-limited peaks in $\phi(\omega,q)$ were reported \cite{Kaplan2025}. The decay of the parametric pump, here due to the Higgs mode's oscillations, with a timescale $\sim \eta^{-1}$ suggests an ``uncertainty relation" for the excited frequencies where $\Delta \omega \sim \eta$; from the dispersion we then obtain $\Delta q = \Delta \omega /v \sim \eta/v$. This is confirmed by analytical calculations for \eqref{eq:second_phi_q} (see Appendix \ref{app:b}), which suggest a Gaussian distribution of $\max_t \phi_q(t) \propto e^{-a(q-q_0)^2}$ with $a\sim v/\eta$ in regime II, independent of $\delta$. Indeed, the peak widths in Fig.~\ref{fig8}(c) are all of the order $\eta/v$. This suggests that the coherence of the spatiotemporal order induced by ultrafast laser pulses is not affected by pulse details, but is only limited by the intrinsic equilibration time of the system with coherence length bound from above by a $(\eta/v)^{-1}$.

\begin{figure}[ht!]
    \centering
    \includegraphics[width=1.0\columnwidth]{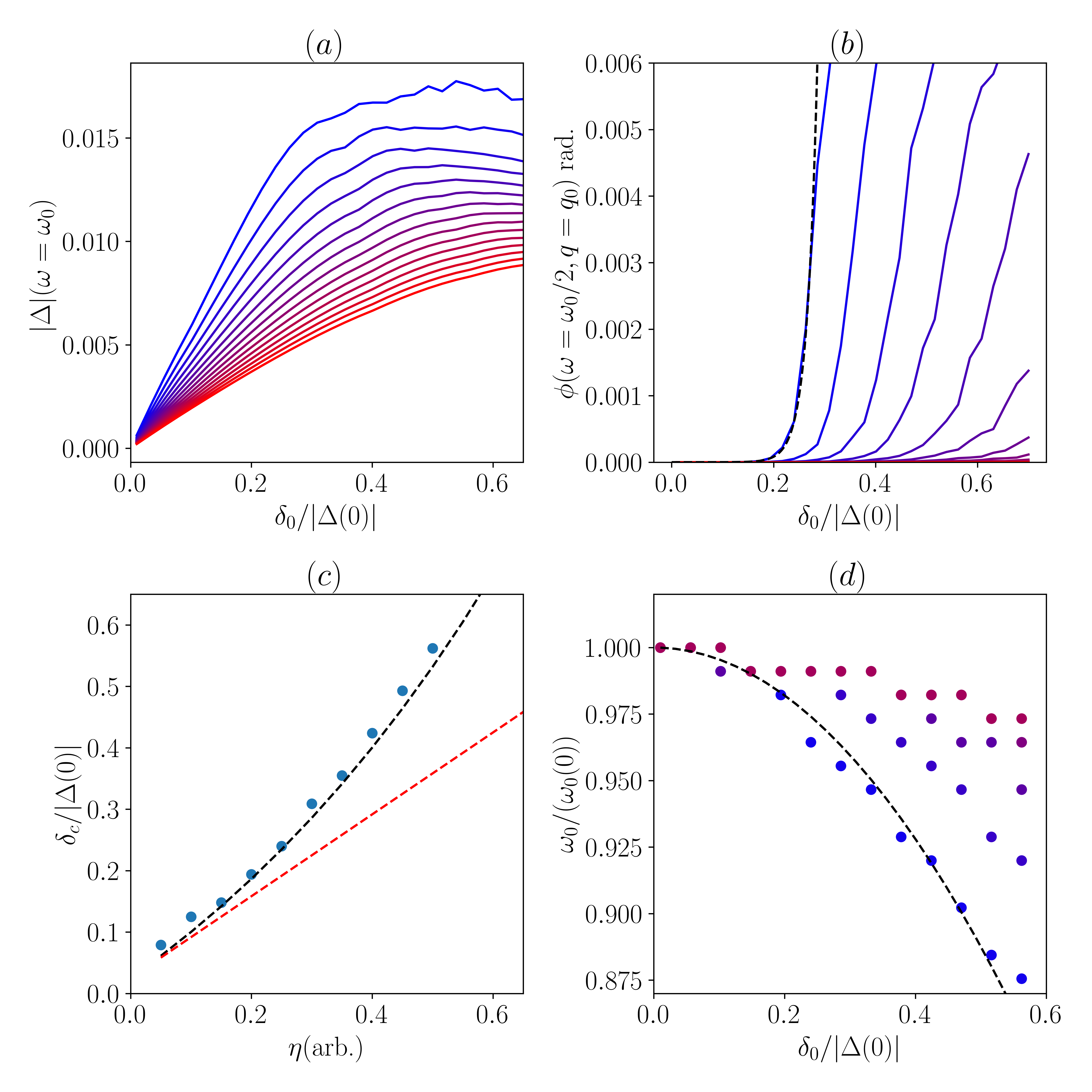}
    \caption{Properties of the Faraday-Goldstone state. (a) Fourier spectrum of $\Delta$ at $\omega=m$, $q=0$, as a function of displacement. The colors -- from blue to red -- indicate increasing dissipation $\eta$. Increasing dissipation leads a low saturation value. (b) Fourier spectrum of $\phi$ at $\omega=m/2$, $q=m/(2v)$ as a function of displacement.  
    A pronounced increase onsets concomitantly with saturation of $\Delta(m,q=0)$.
   The dashed line follows the numerical inversion of Eq.~\eqref{eq:thresh_gam}.
    (c) Displacement threshold vs dissipation. 
    Black dashed line denoted the threshold obtained from an inversion of Eq.~\eqref{eq:thresh_gam}. The red dashed is the equivalent ``continuous driving" threshold \cite{Kaplan2025} $\delta_c \propto \eta$. (d) Frequency softening as a function of displacement. The peak response frequency of the amplitude mode $\omega_{max}/m$ depends on $\delta$ via $\delta^{2}$, approximately, due to intrinsic nonlinearity of the amplitude mode, Eq. \eqref{eq:soft}. Colors (blue to red) indicate increasing dissipation $\eta$.}
    \label{fig3}
\end{figure}

\subsection{Signatures of Faraday-Goldstone Waves at $\mathbf{q}=0$}
Remarkably, the onset of the Faraday-Goldstone wave instability at finite $q_0$ is also clearly observed in the spectroscopy of the Higgs mode at $\mathbf{q}=0$. In Fig.~\ref{fig3}(a), we display the 
spectral weight of
of the amplitude mode at $\omega=m$ as a function
of initial displacement. Initially we observe the expected linear trend. The value of the subsequent saturation  is determined by the dissipation, whose intensity we vary; the resultant peak intensity, as well as the behavior after the onset of parametric oscillations, is plotted in Fig.~\ref{fig3}(a). For weak dissipation the weight of the $\omega=m$ harmonic 
reduces with increasing displacement in the parametric regime, as intensity is redistributed between modes due to nonlinearities and interaction with the $\phi$ mode replicas (secondary peaks in Fig.~\ref{fig2}(d)). 

The onset behavior of the parametric modes is shown in Fig.~\ref{fig3}(b). We plot the amplitude of $\phi(q_0,m/2)$ as a function of initial displacement. A sharp rise is observed to coincide with the saturation of the Higgs mode oscillations, indicating the coherent transfer of energy from the amplitude mode into the phase degree of freedom. A comparison with the linearized analytic theory, Eq. \eqref{eq:aq}, is shown as a dashed black line in Fig.~\ref{fig3}(b).
For large dissipation, the activation of the phase mode is significantly softer, deviating more strongly from the approximate expression (Eq.~\eqref{eq:thresh_gam}).  

The dependence of the threshold on dissipation is plotted in Fig.~\ref{fig3}(c). The linearized equations for phase mode,  Eq. \eqref{eq:aq}, gives a functional form which is obtained by inverting Eq.~\eqref{eq:thresh_gam} (see also App.~\ref{app:b}), and we obtain this for a specific gain, $\textrm{max}_t \phi/\phi(0) =10^3$. In addition, for clarity, we also plot the threshold for the onset of spatiotemporal order for continuous driving \cite{Kaplan2025} (red dashed line, Fig.~\ref{fig2}(c)) which serves as a lower bound for the onset of an instability. 

Intrinsic nonlinearity of the amplitude mode, leading to Higgs' mode frequency shift, Eq.~\eqref{eq:amp}, is also evident in our results. We track the position of the \textit{maximal} intensity $\textrm{max} \lbrace\Delta(\omega)\rbrace$ as a function of displacement. This is shown in Fig.~\ref{fig2}(d). As explained in Sec.~\ref{sec:soft} small deviations of the order parameter in the limit of continuous wave displacement also predict frequency softening. 
Near saturation, for weak dissipation we recover a dependence approximate $\omega/\omega_0 \sim \delta^2$ (Fig.~\ref{fig3}(d), dashed line). For larger dissipation, the frequency softening departs significantly from the expected value, indicating a regime which is both strongly nonlinear and impulsive.  
\subsection{Higgs-Goldstone Beating}
We now describe a key signature of the light-induced nonlinear parametric coupling, which we denote as ``Higgs-Goldstone" oscillations. 
\begin{figure}[ht!]
    \centering
    \includegraphics[width=1.0\columnwidth]{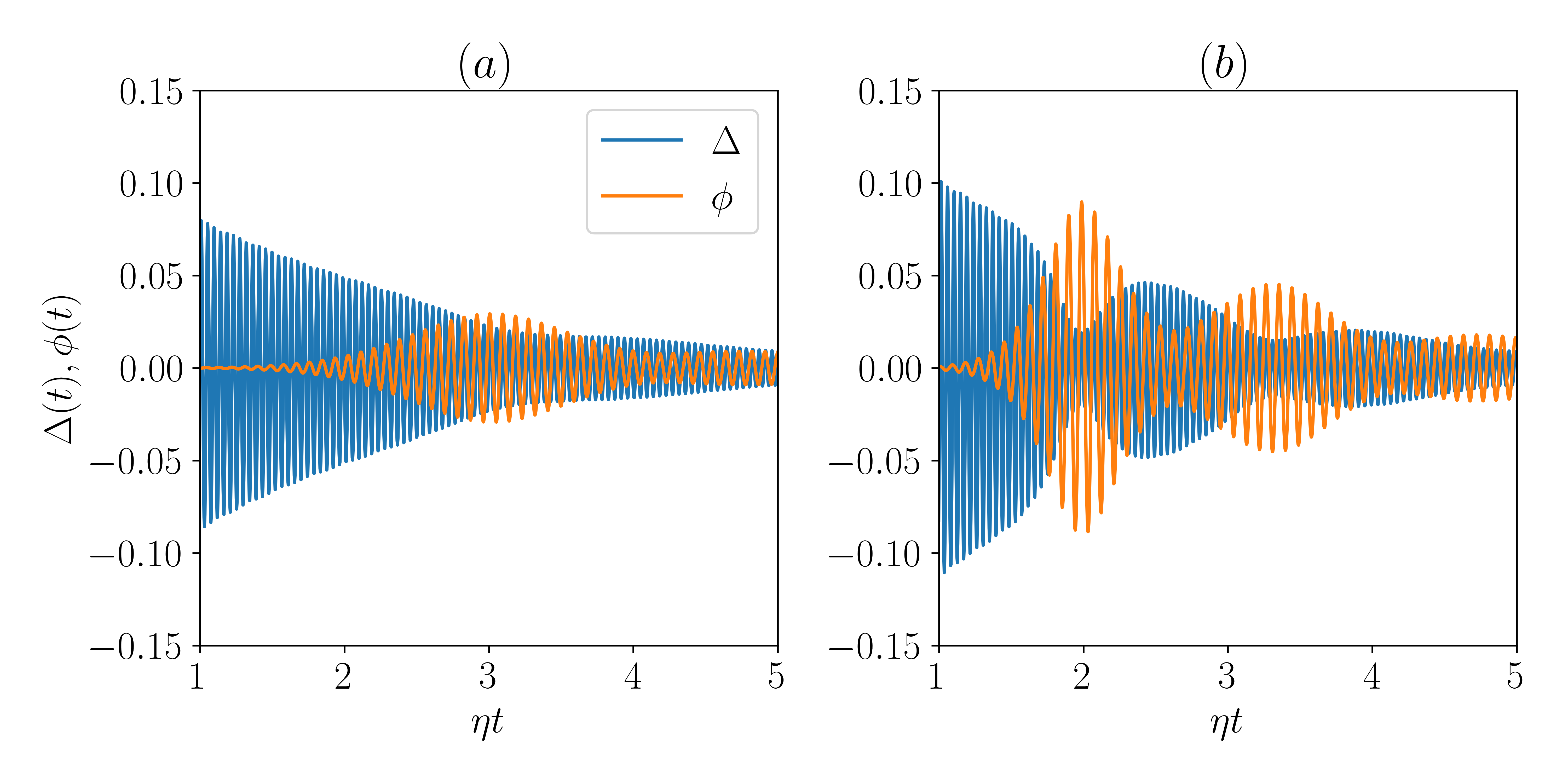}
    \caption{Signatures of Higgs-Goldstone oscillations in time dependence of $\Delta(t,x=0)$ and $\phi(t,x=0)$
    for two values of the drive $\delta$ and small $\eta = 0.07m$. (a) Spatiotemporal order parameter and phase mode oscillations (blue and orange curves), for an initial displacement of $\delta/\Delta(0) = 0.07$.
    (b) Same as (a), but $\delta/\Delta(0) = 0.10$. Here, the amplitude and frequency of beating is directly controlled by the initial displacement $\delta$. Coherent oscillation of spectral intensity between amplitude and phase modes are clearly observed for stronger deviation from equilibrium, shown in (b).}
    \label{fig4}
\end{figure}
For a sufficiently small dissipation $\eta$, the oscillations of the amplitude mode $\Delta(q=0)$ and the Faraday-Goldstone order parameter$\phi(q_0)$ exhibit clearly discernible beating, as we present in Fig.~\ref{fig4}(a)-(b), for two values of the initial pump intensity $\delta$.
The resultant beating is tunable by the drive strength as we contrast in Figs.~\ref{fig4}(a)-(b).

This beating will have clear signatures in the spatiotemporal structure of the order. 
For example, we take a charge density wave (CDW) with nesting vector $q_{CDW}$ (as in, for example, K\textsubscript{0.3}MnO\textsubscript{3}). The charge density is $\rho(x,t) = \textrm{Re}\left(\Delta(x,t)e^{i({q_{CDW}} x+\phi(x,t))}\right)$. Since the amplitude mode does not develop any spatial modulation,
\begin{align}
   \notag \rho(x,t) = \Delta(t)\cos\bigl\{&\phi_0+ q_{CDW}x+ \\ & |\phi_{q,m/2}|\cos(qx+\tilde{\phi})\cos(mt/2)\bigr\}.
   \label{eq:rho_of_x_t}
\end{align}
Here $\phi_0$ is the spontaneously chosen phase of the order parameter in equilibrium; here without loss of generality we take $\phi_0=0$. 
\begin{figure}[ht!]
    \centering
    \includegraphics[width=1.0\columnwidth]{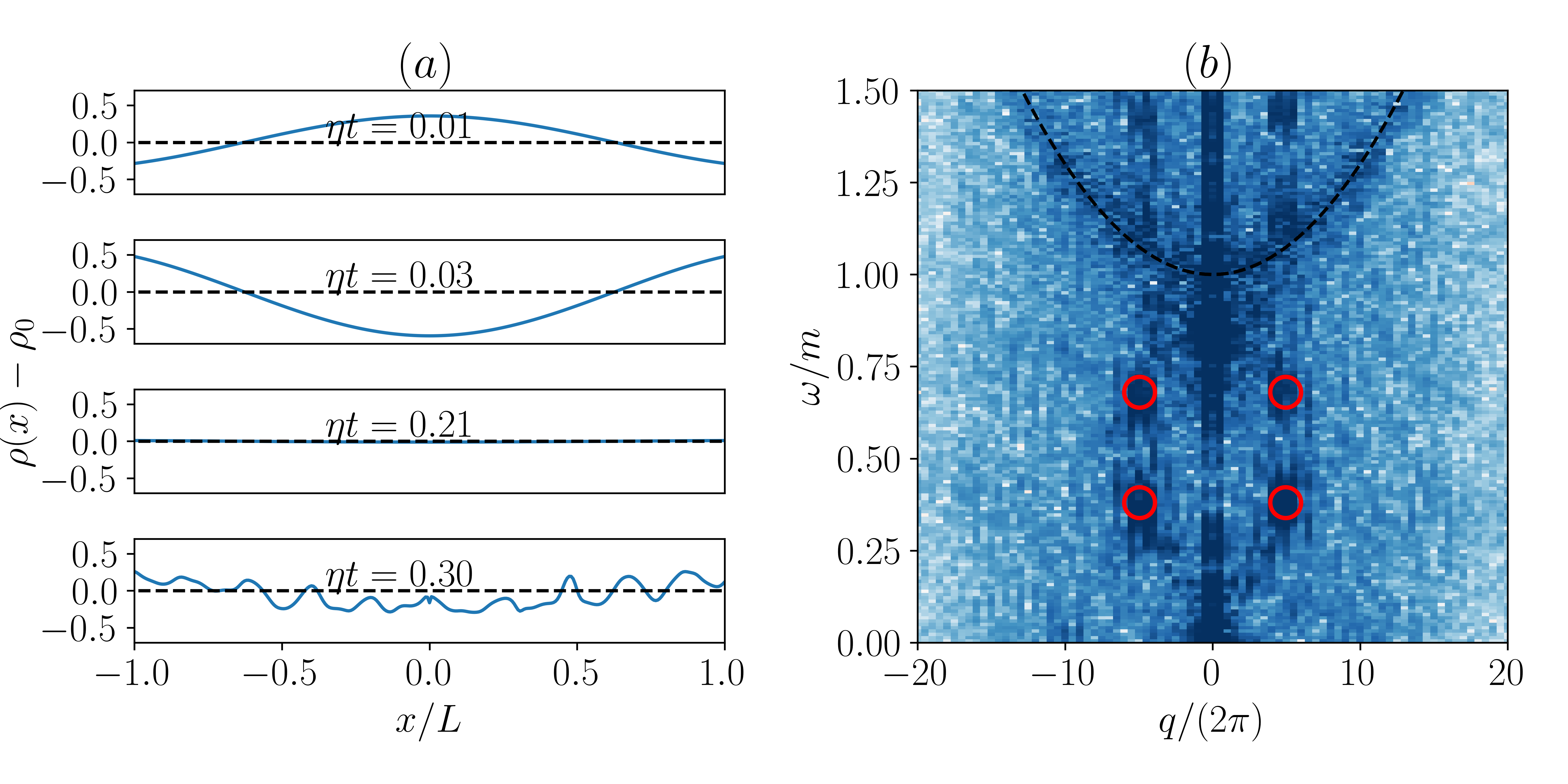}
    \caption{Beating and order in a CDW. (a) Charge density (equilibrium CDW subtracted) as a function space and for different time cuts. Initially, after excitation, orderly oscillations are observed with the node structure of the original CDW nesting vector $q_{CDW}$ still visible. Around $\eta t \approx 0.3$ phase oscillations distort the order significantly, and the original node near $x/L \approx 0.6$ is no longer visible. The resultant structure resembles a ``molten" phase. (b) Spectral function $\rho(q,\omega)$. Despite the seeming lack of order in $(a)$ clear spectral signatures are still discernible: Resonances near $\omega \approx m, q=0$ and parametrically coupled modes near $\omega \approx m/2$. The splitting in frequency reflects the slow beating of the phase mode amplitude, see Fig.~\ref{fig4} arising from nonlinear Higgs-Goldstone coupling.
    }
    \label{fig5}
\end{figure}
Since there is a continuous symmetry-breaking associated with
the development of a Faraday-Goldstone wave, an another spontaneously selected phase $\tilde{\phi}$ appears. 

In Fig.~\ref{fig5}(a) we plot $\rho(x,t)$ in \eqref{eq:rho_of_x_t} after subtracting the equilibrium charge density wave distribution. At short times, before the onset of parametric instability and beating, the equilibrium pattern with nodes at $q_0 x = \pi$ is visible. In this time window, the order parameter merely oscillates at a frequency $\omega \sim m$; it is slightly reduced due to softening as shown in Fig.~\ref{fig3}(d).
After the initial equilibration, a diffuse pattern appears in the charge density due to the phase beating. This is shown in Fig.~\ref{fig5}(a).

We stress, as presented in Fig.~\ref{fig4}(a)-(b), that the amplitude mode itself does not acquire an appreciable spatial modulation; the entirety of the beating comes from coherent oscillations between amplitude and phase. One may therefore  be tempted to describe the state of the matter in Fig.~\ref{fig5}(a) as a ``molten" CDW. However, a spectroscopic analysis of this state in Fig.~\ref{fig5}(b) reveals the structure of the ordered phase is mostly intact. We find the dispersion of the amplitude mode (dashed line, Fig.~\ref{fig5}(b)) and its central frequency 
still present; the latter is softened due to nonlinear interactions. Moreover the peak at $m/2$ is split into two satellites, characteristic of the slow beating of the amplitude of the Goldstone mode. Thus this phase combines strong signatures of Higgs-Goldstone coupling, parametric down-conversion and beating phenomena.

\subsection{Finite temperature}
\label{sec:temp}
So far, we have presented results at $T=0$ and we must consider 
$T \neq 0$ if we want to make contact with experiment.  Finite temperature would affect the results in two ways. First, the Higgs mass, $m = \frac{2\sqrt{\alpha}}{2v}$ is renormalized to 
$\alpha = \alpha_0(1-T/T_c)$. This change will be reflected in the threshold and the ordering wavevector $q$ which both depend $m$. Working deep in the ordered phase ($T/T_c \ll 1$), we can safely neglect these changes and focus instead on the properties of the phase degree of freedom.  The emergent coupling between the Higgs and the Goldstone modes depends on the quartic coefficient of the Landau theory. For the present discussion, we neglect the thermal softening of the Higgs mass.
At finite $T$ the massless Goldstone mode is expected to execute a random walk \cite{campa2009statistical}. In the present 1D example considered in Sec.~\ref{sec:onset}, the phase fluctuations are expected to be severe due to the Mermin-Wagner theorem \cite{mermin1966absence,hohenberg1967existence}.
{\allowdisplaybreaks
\begin{figure}[ht!]
    \centering
    \includegraphics[width=1.0\linewidth]{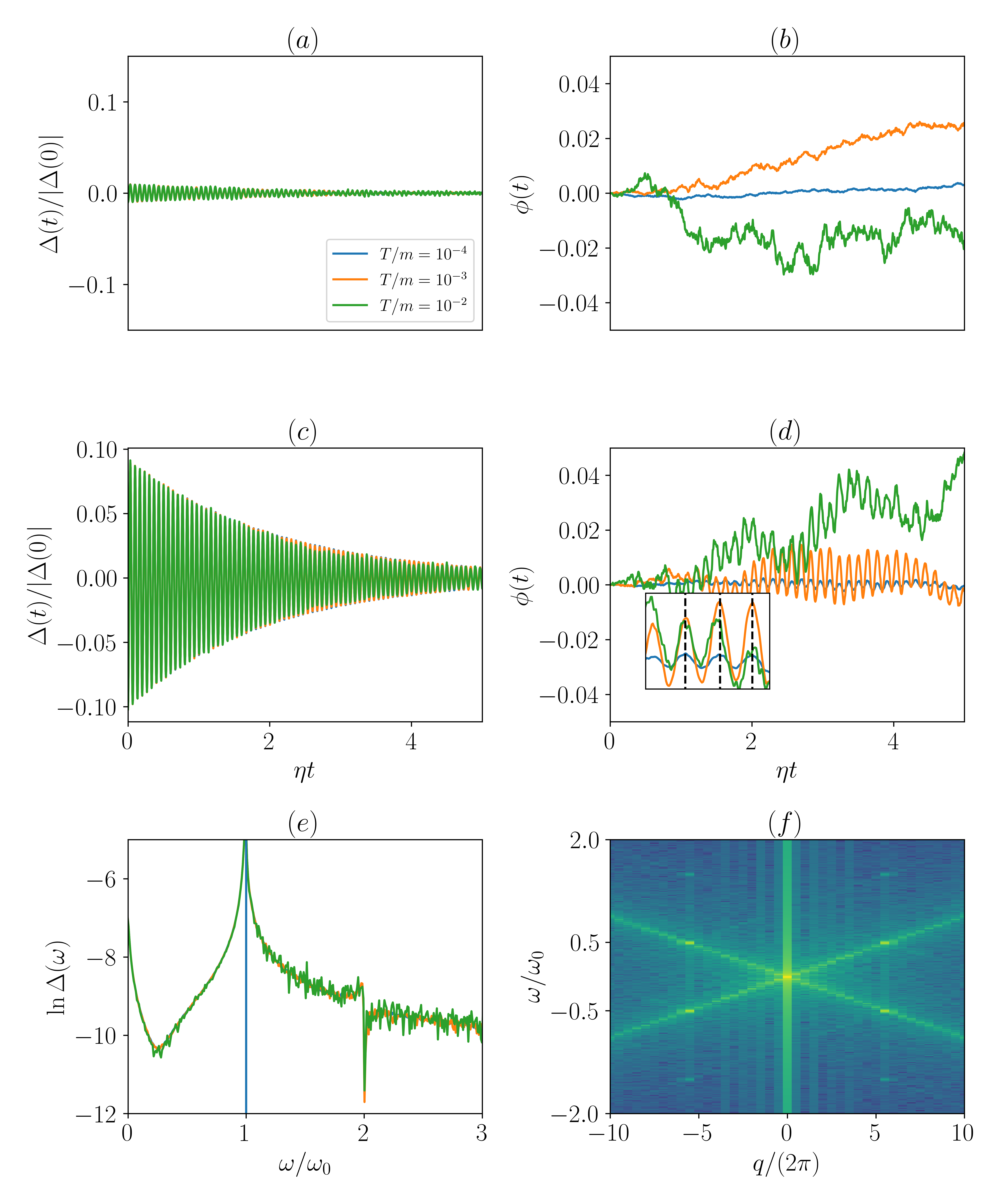}
    \caption{Temperature dependent properties of the Higgs-Goldstone coupling. (a) Amplitude mode oscillations at three select temperatures. The amplitude mode oscillates without substantial impact from temperature as $T \ll m$. (b) Phase mode $\phi(0,t)$ oscillations \textit{below} the threshold. The phase show random drift similar to Mermin-Wagner fluctuations, due its massless character. (c) Above the threshold, the amplitude mode oscillates similar to the $T=0$ results. (d) The phase mode shows both the coherent oscillations at $m/2$ (Fig.~\ref{fig2}(b)) with the additional drift of the overall phase of the underlying order; inset shows a zoomed-in image of the dynamics of the phase at different temperatures, revealing that the phase of the induced order does not drift. (e) Log of the spectral function of $\Delta$ showing a clear resonance at $\omega=m$ and thermally activated incoherent modes. (f) Log of the spectral function of $\phi(x,t)$. The peaks corresponding to the underlying order are clearly visible, even at the highest temperature $T=10^{-2}m$. The drift of the parent state's phase is seen as the large peak at $\omega =0,q=0$ (contrast with Fig.~\ref{fig2}(d)). }
    \label{fig7}
\end{figure}
Indeed, below the threshold, the phase carries out the expected random walk, causing the underlying order to drift as shown in Fig.~\ref{fig7}(b). The amplitude mode (Fig.~\ref{fig7}(a)) displays regular oscillatory motion with little thermal effects due to its gapped nature. As expected \cite{kardar2007statistical}, the correlation $\langle \phi(0)\phi(t)\rangle \sim t$, is clearly visible in Fig.~\ref{fig7}(b) as a linear drift in time of the phase.

In the parametrically excited regime (Fig.~\ref{fig7}(c)-(d)), we find similar dynamics for $\Delta$ (Fig.~\ref{fig7}(c)) as in Fig.~\ref{fig2}(a), but sharply different ones for the phase $\phi(x,t)$ (Fig.~\ref{fig7}(d)). In addition to clear signatures of parametrically excited oscillations, an overall thermal drift of the original phase is present. Remarkably, however, the phase of parametrically driven order is \textit{not} thermally affected. We show this by plotting $\phi(x_0,t), ~ x_{0}=\pi/q_0$, for various temperatures (Fig.~\ref{fig7}(d)-inset). Unlike the expectation from a thermally drifting order, the maxima and minima line up for all time-slices. This is conclusive evidence that the Faraday-Goldstone waves are more robust to effects of temperature than an equilibrium symmetry-breaking order, suggesting a tantalizing possibility of inducing spatiotemporal order in a thermally disordered state.}

\section{Comparison with Experiment}
{\allowdisplaybreaks
We are now in a position to test the applicability of the theory in experimental settings. To this end, we take the prototypical 1D CDW material, \bbronze \cite{pouget1985structural,schutte1993incommensurately,schlenker2012low,Gruner_CDW}. 
As a prototypical host realizing a Peierls phase \cite{frohlich1954theory}, it has been extensively studied via pump-probe measurements \cite{tomeljak2009dynamics,schaefer2014collective,schafer2010disentanglement,Liu13,Mankowsky2017}. In such experiments, the system is irradiated with an excitation (pump) and the effect of the latter is measured through an optically accessible quantity such as reflectivity. While our theory cannot directly obtain quantitative estimates for the change in electronic properties (to be explored in future work), we consider $\Delta R(\omega)/R$ where $\Delta R$ is the change in reflectivity at the frequency $\omega$, to be a measure of the amplitude mode variation.

The variable parameters are the fluence $I$ of incoming photons in $mJ cm^{-2}$
 and the central frequency of the pulse. Here, we consider an excitation at $\lambda= 6.5 \mu m$. We use the relative change in reflectivity to track two key observables: the maximum amplitude of the displacement, measured through the maximal shift in reflectivity; and the peak frequency of the amplitude mode oscillation %maximal oscillation 
 -- the renormalized $\omega_0$ in our theory. Crucially, we require as few fitting parameters as possible. We find the following minimal recipe affords excellent agreement with experiment: a single rescaling of the experimental data by $\Delta R \to \Delta R/115$ and $ I \to I/39$. This establishes the interconversion: $1/30 (mJ cm^{-2}) =  \frac{\delta_0}{|\Delta(0)|}$ (displacement). The theory includes one free parameter- dissipation $\eta$ ; this controls the saturation value and onset of parametric excitation; this is the last value to be determined. 
\begin{figure}[ht!]
    \centering
    \includegraphics[width=1.0\columnwidth]{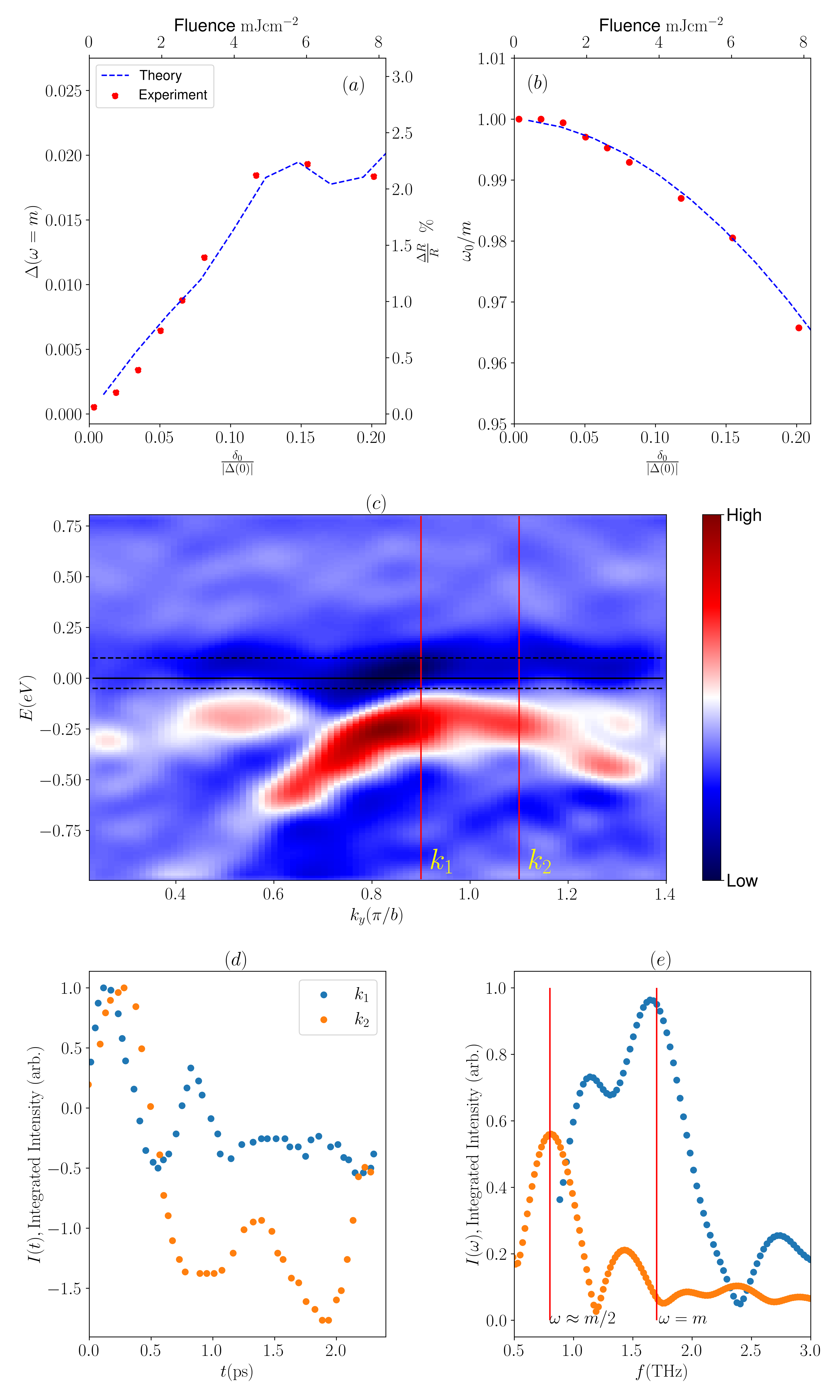}
    \caption{Comparison with experiment and theory on pump-probe experiments \bbronze. (a) Comparison of amplitude mode displacement (left axis) and relative reflectivity change (right axis) as a function of fluence (top axis) and amplitude mode displacement (bottom axis). (b) Frequency mode softening as measured experimentally vs the predicted signature. The single-scaling function as explained in the main text is derived from this comparison between theory and experiment. The dissipation is then adjusted to match the onset of the plateau in $\Delta(\omega)$. Data adapted from Ref.~\cite{Mankowsky2017}. (c) ARPES data on \bbronze in equilibrium, showing $\frac{\partial^2 I}{\partial E^2}$ for greater clarity. Two momenta $k_1, k_2$ are highlighted corresponds to the regions most strongly activated after illumination. (d) Integrated intensity (black dashed line range in (c)) as a function of delay from the pump for the $k_1$ and $k_2$ points. (d) Fourier spectrum of (d) showing clear resonance at $m$ for $k_1$ but at $m/2$ for $k_2$. Data and details are found in Ref.~\cite{Liu13}.}
    \label{fig6}
\end{figure}
We find excellent agreement for the value of $\eta=0.15m$. In Fig.~\ref{fig6} we plot these  observables with the above rescaling, against two $x,y$ axes: one describing the parameters of the theory $\Delta,\omega_0$ vs the displacement and fluence, respectively. In Fig.~\ref{fig6}(a) we find overall good agreement between experiment (dots) and theory (solid line). For small fluences, experimental data contain an upturn not captured by our theory, likely due to the presence of other, competing couplings. Our theory further predicts fine structure not accessible in experiment due to the lack of data points. In Fig.~\ref{fig6}(b) we plot the frequency softening predicted by theory for the same $\eta$ and rescaling extracted from the data of Fig.~\ref{fig6}(a). Here, the agreement is superb, and indicates that the dominant processes contributing to the frequency variation are well-captured by the theory. 

Using ARPES data from Ref. \cite{Liu13}, plotted in Fig.~\ref{fig6}(c), we examine the effect of the presence of spatiotemporal order on the electronic states of \bbronze. In Fig.~\ref{fig6}(c) we plot $\frac{\partial^2 I}{\partial E^2}$ focusing on regions where strong intensity variations are observed after impulsive excitation (see details in Ref.~\cite{Liu13}). We select two points in momentum space $k_1,k_2$ superimposed on Fig.~\ref{fig6}(c). $k_1$ corresponds to the original gap opening position in momentum space, and $k_2$ is from a proximate bands that undergoes level repulsion with the folded state. We then plot the intensity integrated over a small energy window, between $-50$ and $+100$ meV (horizontal dashed lines in Fig.~\ref{fig6}(c) as a function of delay $t$ in Fig.~\ref{fig6}(d) for both $k_1, k_2$. The temporal Fourier transform of the intensity at $k_1$ and $k_2$, Fig.~\ref{fig6}(e), reveals a striking contrast. We find a sharp peak at $f_1 = 1.7 \textrm{THz}$ which corresponds to the well-established amplitude mode frequency in \bbronze (Refs.\cite{schafer2010disentanglement,schaefer2014collective}). We identify this frequency with the mass the amplitude mode in our theory, $m$. No additional peaks are seen at half the frequency for $k_1$. In contrast, $k_2$ shows that its dominant oscillation is derived from $m/2$ oscillation, with the central frequency $\omega \approx 0.8\textrm{THz}$. Clearly $k_2$ contains the parametric oscillation derived from the phase mode frequency. The electronic states are affected by a potential of the form $V_q \sim V \cos(q_0x+\bar{\varphi})\cos(mt/2)$, and its intensity is largest near avoided crossings of bands, as seen for $k_2$. This is further signature of the existence of parametrically down-converted modes in the system.  
We speculate that beating has already been observed as well, in Ref.~\cite{Mankowsky2017} (Fig. 3(a), therein). Around $t \sim 2\textrm{ps}$ of delay, dips in the intensity of reflectivity variation are strongly reminiscent of the evanescence of the amplitude mode in Fig.~\ref{fig4}(a), which corresponds to the shift of oscillation weight to the phase mode. We expect future experiments to resolve the beating more accurately, and compare it against our predictions. }

\section{Discussion}

Using a combined numerical and analytic approach, we have studied parametric instabilities driven by light in systems with broken continuous symmetries. The standing waves at
finite momentum that we have presented are precisely the solid-state analogues of Faraday waves that are found in driven liquids and suspensions
\cite{Faraday1831,Cross93,Nguyen19,Liebster25}. Since the pattern formation occurs via the massless mode of the symmetry-broken state, we call this phenomenon Faraday-Goldstone waves -- which, to our knowledge, has not been discussed before.
There is a threshold to the Faraday wave formation, beyond which the Goldstone modes at $\pm q_0$ and $m/2$ become coherent with several observable consequences.
 
For sufficiently weak dissipation we have shown
there will be Higgs-Goldstone beating associated with
the coherent transfer of energy between the two modes;
this is a zero-momentum signature of the onset of parametric instability. 

Including finite temperature fluctuations (Sec.~\ref{sec:temp}), we observed that the phase of the Faraday-Goldstone waves weakly couples to an external bath, even though the global phase in our one-dimensional calculation strongly drifts with temperature. 
Importantly, this demonstrates that quasi long-range may be permitted in the driven system, even when the equilibrium state cannot order.

Comparing our theoretical results with those from pump-probe measurements on \bbronze, we have found excellent agreement for a wide range of fluence, Fig.~\ref{fig6}(a)-(b). 

The experimental data contains two distinct and independent checks of our theory: the amplitude mode saturation (Fig.~\ref{fig3})
and the frequency softening (Fig.~\ref{fig6}. 
In both theory and experiment, we emphasize that this frequency softening is inconsistent with melting that was only observed close to $T_c$ \cite{schafer2010disentanglement}; sufficiently deep in the ordered state, the Higgs-Goldstone regime is the likely source of softening and nonlinearities.

The consequences of the coherent energy exchange mechanism studied here are profound: we suggest that even in the absence of long-range \textit{static} ordering, quasi-long-range order \textit{can} can persist dynamically, at least on the scale of the correlation length \cite{halperin2019hohenberg} which will be set by the width of the resonance at $m/2$ and $q_0$. This suggests the possibility of optically exploring other kinds of non-thermal behavior such as Kibble-Zurek physics  \cite{zurek1985cosmological}, time-crystalline order \cite{Zaletel23,Wilczek2012,Sacha_2017,Morrell25} and many-body localization \cite{altman2018many}. Our approach is also amenable to first-principle calculations \cite{kaplan2025spatiotemporalorderparametricinstabilities}
for the identification of specific material hosts. 
More generally our results offer a new pathway towards 
the design of  
non-thermal incommensurate patterns 
in strongly interacting quantum materials with optical pulses.
\begin{acknowledgments}
We thank B. Doucot, G. Fiete, N. Gedik, D. Juraschek, 
S. Sachdev and T. Senthil for inspiring discussions. We are indebted to V. Oganesyan for brainstorming with us in the very
early stages of this work. 
We also acknowledge A. Kogar for illuminating conversations both
on this project and on a related collaboration. DK is supported by the Abrahams Postdoctoral Fellowship of the Center for Materials Theory, Rutgers University and the Zuckerman STEM Fellowship. DK, PAV and PC acknowledge the hospitality of
the Aspen Center for Physics, which is supported by National Science Foundation grant PHY-2210452.
PAV acknowledges support by the
University of Connecticut OVPR Quantum CT seed grant.
\end{acknowledgments}
\clearpage

\appendix
\section{Equations of motion for phase and amplitude}
\label{App:A0}
The equations of motion for $\Psi(x,t)$ obtained from $\frac{\delta \mathcal{F}}{\delta \Psi^{*}} = -\eta \partial_t \Psi$ are:
\begin{equation}
    \partial_t^2 \Psi+\eta \partial_t \Psi - v^2 \partial_x^2 \Psi+ \alpha \Psi+ \beta |\Psi|^2 \Psi = [\mathcal{R}_1(x,t) +i \mathcal{R}_2(x,t)],
    \label{eq:app:eom}
\end{equation}
where the right-hand side represents thermal noise with real and imaginary part uncorrelated, i. e. $\langle \mathcal{R}_i(x,t) \mathcal{R}_j(x',t') \rangle = \eta T \delta_{ij} \delta(x-x')\delta(t-t')$. The equations for phase $\phi$ and amplitude $\Delta$ are obtained by introducing $\Psi(x,t) = |\Delta(x,t)|e^{i\phi(x,t)}$ to Eq. Eq.~\eqref{eq:app:eom}, multiplying the equation by $e^{-i\phi(x,t)}$ and then taking the real and imaginary parts of the equation:
\begin{equation}
    \begin{gathered}
        \left[-v^2 \partial_x^2+\partial_t^2+\eta\partial_t \right]|\Delta| + |\Delta|(\alpha + D(\phi)) + \beta|\Delta|^3 = \mathcal{R}_\Delta(x,t) .
        \\
        \Delta (\partial_t^2-v^2 \partial_x^2) \phi +2 (\partial_t \Delta \partial_t \phi - v^2 \partial_x \Delta \partial_x \phi)+\eta \Delta \partial_t \phi = \mathcal{R}_\phi(x,t) 
    \end{gathered}
    \label{eq:app:eom2}
\end{equation}
where 
\begin{equation}
    \begin{gathered}
\mathcal{R}_\Delta(x,t)  =  {\rm Re}\left([\mathcal{R}_1+i \mathcal{R}_2]e^{-i\phi}\right),
        \\
\mathcal{R}_\phi(x,t)  =  {\rm Im}\left([\mathcal{R}_1+i \mathcal{R}_2]e^{-i\phi}\right).
    \end{gathered}
\end{equation}
Evaluating the averages one finds $\langle \mathcal{R}_\alpha \mathcal{R}_\beta \rangle = T\eta \delta_{\alpha\beta} T\delta(x-x')\delta(t-t')$. Multiplying the second equation in \eqref{eq:app:eom2} by $\Delta$, one obtains \eqref{eq:phase_eq} and \eqref{eq:amp_eq} of the main text.

\section{Numerical details}
\label{app:num}

In this section, we detail aspects of the numerical evaluation of Eqs.~\eqref{eq:phase_eq}-\eqref{eq:amp_eq}. The spatial grid $x$ is discretized in one-dimension with $N_x = 501$ points. The Laplace operator is constructed by taking the fields $\phi,\Delta$ as $\phi(x) \to \phi(x_n)$ and $\partial_x^2 \phi \approx \frac{\phi(x_{n+1})-2\phi(x_n)+\phi(x_{n-1})}{\Delta x^2}$ and we impose periodic boundary conditions, such that $\phi(x_{N_x+1}) = \phi(x_1)$. Temporal integration is carried by converting Eqs.~\eqref{eq:phase_eq}-\eqref{eq:amp_eq} into coupled, first-order in time differential equations, that will also permit adding a thermal bath \cite{leimkuhler2013robust}. We let $v_\phi = \partial_t\phi$, and thus the equations are $\dot{v}_\phi = V(\phi,\Delta)$. Dissipation is included through a half time-step integration,
$v_{\phi}({t+\Delta t/2}) = e^{-\eta \Delta t/2}v_{\phi}(t)+\mathcal{W}\sqrt{1-e^{-\eta \Delta t}}$, where $\mathcal{W}$ is a normal distribution whose correlation function is set by fluctuation-dissipation and is equal to $\eta T$. The discretization of both $\phi, v_\phi$ in time is along $N_t = 3000$ points, with the sampling rate determined so as to accommodate $2.5 m$ where $m$ is the Higgs mass. Total integration time used as $t_{\textrm{all}} = 10^6\eta$. We parametrize the problem in terms of the original quartic Higgs potential. Thus, in Figs.~\ref{fig1}-\ref{fig5} and Fig.\ref{fig7}, we used $\alpha=-100$, $\beta=0.1 |\alpha|$. $\eta$ values are found in the figure captions, where appropriate. In Fig.~\ref{fig8}, for better momentum space resolution, we used $\alpha=-600$, $\beta=0.1 |\alpha|$. The fits done in Fig.~\ref{fig6} were carried when the model was solved for $\eta = 0.15$, $\alpha=-600$, $\beta=0.1|\alpha|$. No additional tuning was carried out. 

All initial conditions were averaged over, in the following way: $N_r = 30$ realizations of the random initialization of the field, which was carried out using a uniform distribution in space $|A|\mathcal{U}(-1,1)$, where $|A|$ is the maximum possible value at any position $x$. The spectral functions $|\phi(q,\omega)|$ were obtained by averaging $\tilde{\phi}_{q,\omega} = \frac{1}{N_r} \sum_n |\phi^{(n)}(q,\omega)|$.

\section{Parametric instabilities with decaying drive and detuning}
\label{app:b}
Here we generalize the standard derivation of parametric instabilities \cite{landau2013mechanics} to the case of a decaying parametric drive. Our starting point is Eq. \eqref{eq:second_phi_q}, which we generalize to the case of an arbitrary decay constant of the drive, i. e.
\begin{equation}
     \ddot{\phi}_q+\eta\dot{\phi}_q +v^2q^2\phi_q -m\delta e^{-\lambda t}\sin(mt)\dot{\phi_q} = 0.
     \label{app:eqparam}
\end{equation}
As this is a linear equation, we can solve separately for ${\rm Re} \phi_q$ and ${\rm Im} \phi_q$; without loss of generality we focus on the former. We seek the solution in the following form:
\begin{equation}
{\rm Re} {\phi}_q(t) = a_q(t) e^{i m t/2} + a_q^*(t) e^{-i m t/2},
\end{equation}
where $\dot a_q/a_q \ll m$ is expected. The Eq. \eqref{app:eqparam} takes then the form:
\begin{equation}
\begin{gathered}
e^{i m t/2}\biggl( (\eta+i m )\dot a_q +(v^2 q^2 - m^2/4+i \eta m/2 ) a_q +
\\
+\frac{m}{4} \delta e^{-\lambda t} (m a_q^*+2 i \dot a_q^*)
\biggr) + c.c. + ... = 0,
\end{gathered}
\end{equation}
where $...$ contains smaller terms oscillating with frequency $m$ or larger in absolute value as well as terms containing higher derivatives of $a$. As the typical frequencies in the Fourier transform of $a_q(t)$ are much less than $m$, the expression inside the brackets has to vanish identically. Moreover, using $\eta\ll m$ and $\dot a_q/a_q \ll m$ we further approximate the resulting equations, yielding:
\begin{equation}
    im \dot a_q +  i\frac{m\delta}{2} \dot a_q^*  = - (m \varepsilon + i \eta m /2) a_q + \frac{m\delta}{4} e^{-\lambda t} a_q^*=0,
\end{equation}
where we have introduced the notation:
\[
\varepsilon = (v^2 q^2-m^2/4)/m
\]
As will be shown below, the relevant values of $m$ are of larger but not too much large than $\gtrsim \eta/m \ll 1$ allowing to neglect $m\delta$ compared to $m$. The final approximate system of equations for $a_q(t),a_q^*(t)$ is most conveniently written in a matrix form:
\begin{equation}
\frac{d}{dt}
\begin{pmatrix}
a_1\\
a_2
\end{pmatrix}
    = \begin{pmatrix} 
    - \eta/2 & d e^{-\lambda t} +\varepsilon \\
d e^{-\lambda t} -\varepsilon&  - \eta/2 
    \end{pmatrix}
     \begin{pmatrix}
a_1\\
a_2
\end{pmatrix},
\label{eq:app:matr}
\end{equation}
where $d\equiv\frac{m\delta}{4}$, $a_1 = {\rm Re} a_q$, $a_2 = {\rm Im} a_q$. It can be appreciated that the matrix in r.h.s. of Eq.~\eqref{eq:app:matr} does not commute with itself at a different time. Therefore, a general solution can be only written as a time-ordered exponential which complicates the analysis in a general case. Consequently, below we follow a different approach:
\\
1. We solve the equations exactly for $\varepsilon = 0$ (where this issue does not arise);
2. We obtain an approximate solution for $\lambda \ll \eta$. Remarkably, the obtained solution also reproduces the exact result in $\varepsilon\to 0$ limit.
3. We use this approximate solution to provide qualitative estimates for the actual problem with $\lambda= \eta/2$.
\subsection{Resonant case $vq=m/2$}
In this case, the solution of Eq. \eqref{eq:app:matr} takes a simple form:
\begin{equation}
    \begin{pmatrix}
a_1\\
a_2
\end{pmatrix} (t)  = 
\exp\left(-\frac{\eta t}{2} + d\frac{1-e^{-\lambda t}}{\lambda} \tau_x \right) 
\begin{pmatrix}
a_1\\
a_2
\end{pmatrix}(0)
\label{eq:app:solres}
\end{equation}
where $\tau_i$ are Pauli matrices. From this one can find both the rate of growth of the solution with $\delta$ as well as any other observables. Specifically, growth is achieved for $\tau_x\to +1$.
\\
For example, one can calculate the Fourier transform 
\begin{equation}
\begin{gathered}
    L = \left| \int_0^\infty dt' x(t) e^{i \omega t} \right| \approx \int_0^\infty dt' e^{-\frac{\eta t'}{2}+ \frac{d(1 - e^{-\lambda t'})}{\lambda}} x_+(0) =
    \\
    =\frac{1}{\lambda} e^{ \frac{d}{\lambda}} \left(\frac{d}{\lambda}\right)^{-\frac{\eta}{2\lambda}}
    \left(
    \Gamma\left[\frac{\eta}{2\lambda}\right]
    -
    \Gamma\left[\frac{\eta}{2\lambda}, \frac{d}{\lambda}\right]
    \right) x_+(0),
\end{gathered}
\end{equation}
where $x_+(0)$ is the initial condition for the symmetric ($a_1=a_2$) solution. In the limit $\lambda\to 0$ the integral simply yields $\frac{1}{\eta/2- d}$ resulting in the conventional condition for parametric amplification onset.

For $\lambda =\eta/2$ realized in the main text, the amplification compared to the no driving case can be computed as follows (neglecting the $\tau_x\to-1$ solution):
\begin{equation}
    \frac{L(m \delta/2 \eta \equiv z) }{L(0)}
    =
    \frac{e^{z}}{z} 
    \left(
   1
    -
    \Gamma\left[1, z\right]
    \right).
    \label{eq:thresh_gam}
\end{equation}

It is also straightforward to find the maximal value of $x(t)$ by maximizing Eq. \eqref{eq:app:solres}, neglecting the decaying $\tau_x=-1$ solution. The maximum is achieved for $t= \frac{1}{\lambda} \log \frac{2d}{\eta}$ and is equal to (taking the $\lambda=\eta/2$ case):
\begin{equation}
    \frac{a_{max}}{a_0} = \exp\left(\frac{m \delta }{2\eta} - 1 -  \log \frac{m \delta } {2\eta} \right).
\end{equation}

\subsection{Nonresonant case in the adiabatic limit $\lambda\ll \eta$}
We begin with the $\lambda = 0$ case which has the exact solution:
\begin{equation}
\begin{gathered}
    \begin{pmatrix}
a_1\\
a_2
\end{pmatrix} (t)  = 
e^{-\frac{\eta t}{2} }
\left[
\cosh\left(\sqrt{d^2-\varepsilon^2} t\right) \right.
\\
\left.
+
\frac{d \tau_x + i \varepsilon \tau_2}{\sqrt{d^2-\varepsilon^2}}
\sinh\left(\sqrt{d^2-\varepsilon^2} t\right)
\right] 
\begin{pmatrix}
a_1\\
a_2
\end{pmatrix}(0).
\end{gathered}
\label{eq:app:sollam0}
\end{equation}
For long times $t \gg 1/\sqrt{d^2-\varepsilon^2}$ one observes that the exponential parametric enhancement onsets at $d>\sqrt{(\eta/2)^2+\varepsilon^2}$. Moreover, the solutions for different $\varepsilon$ will exhibit different growth rate $\propto e^{(\sqrt{d^2-\varepsilon^2} -\eta/2)t}$ (the leading corrections to this arise from the matrix structure of \eqref{eq:app:sollam0} and are of the order $\varepsilon/d$). For a fixed time $t$ we can thus define a full width at half maximum for the solution magnitude as a function of $\varepsilon$ from $e^{(\sqrt{d^2-\varepsilon^2} -\eta/2)t} = e^{(d -\eta/2)t}/2$:
\begin{equation}
    \Delta \varepsilon^{FWHM}_{\lambda=0} 
    =
    2\sqrt{\frac{\log 2 (2 dt -\log 2)}{t^2}}.
\end{equation}

We now generalize this solution to the case of finite $\lambda\ll \eta$. To do this, we first transform Eq. \eqref{eq:app:matr}. First, we introduce:
\begin{equation}
         \begin{pmatrix}
a_1\\
a_2
\end{pmatrix}(t)
\equiv
e^{-\eta t/2}
e^{f(t) \tau_x}
         \begin{pmatrix}
a_1'\\
a_2'
\end{pmatrix}(0),
\label{eq:app:aprime}
\end{equation}
where $f(t) = d \frac{1-e^{-\lambda t}}{\lambda}$. Then, introducing $a'_\pm(t) = (a_1'\pm a_2')/\sqrt{2}$ we can find the equations for $a'_\pm(t)$ to be:
\begin{equation}
    \begin{gathered}
        \dot a_+' = - \varepsilon e^{-2 f(t)} a_-',
        \\
        \dot a_-' = \varepsilon e^{2 f(t)} a_+'.
    \end{gathered}
    \label{eq:app:apam}
\end{equation}

For $\varepsilon = 0$, $a_\pm'(t) = {\rm const}$ and thus we recover \eqref{eq:app:solres}. In general case, we can reduce the complexity of Eq. \eqref{eq:app:apam} as follows. We divide the first equation by $a_+'$ and the second one by $a_-'$ and then subtract them, which yields a single equation for $a_+'/a_-'$. Introducing $z\equiv e^{2 f(t)} a_+'/a_-'$ we finally get an equation:

\begin{equation}
\begin{gathered}
    \dot z = -\varepsilon z^2 +2d(t) z - \varepsilon =
    \\
    =-\varepsilon\left(z-\frac{d(t)+\sqrt{d^2(t)-\varepsilon^2}}{\varepsilon}\right)
    \left(z-\frac{d(t)-\sqrt{d^2(t)-\varepsilon^2}}{\varepsilon}\right),
\end{gathered}
\end{equation}
where we introduced a shorthand notation $d(t) = d e^{-\lambda t}$. The r.h.s. of this equation has two fixed points as long as $d(t)>\varepsilon$, the attractive is the larger one $z =\frac{d(t)+\sqrt{d^2(t)-\varepsilon^2}}{\varepsilon}$. This implies that as long as $d(t)$ is a sufficiently slow function (adiabatic approximation) $z$ will simply track the fixed point at long times $t\gg \frac{1}{d(0)} \log[2d(0)/\varepsilon]$, assuming $d(0)\gg \varepsilon$ (to be demonstrated below). More precisely, this condition is $\lambda\ll \sqrt{d^2(t)-\varepsilon^2}$, which, as we will see below is satisfied for $\lambda \ll \eta$ for sufficiently long times, such as to capture the maximal value of $a(t)$.

Under the adiabatic approximation one then gets the following for $a_\pm'$:
\begin{equation}
    \begin{gathered}
        \dot a_+' \approx - \left(d(t)-\sqrt{d^2(t)-\varepsilon^2}\right) a_+',
        \\
        \dot a_-' \approx \left(d(t)+\sqrt{d^2(t)-\varepsilon^2}\right) a_-'.
    \end{gathered}
    \label{eq:app:apam_adiab}
\end{equation}
Using the definition \eqref{eq:app:aprime} we finally get:
\begin{equation}
\begin{gathered}
    a_+(t) \approx e^{-\eta t/2 } e^{\int_0^t dt' \sqrt{d^2(t)-\varepsilon^2} } a_+(0),
    \\
    a_-(t) \approx e^{-\eta t/2 } e^{\int_0^t dt' \sqrt{d^2(t)-\varepsilon^2} }a_-(0),
\end{gathered}
\end{equation}
where $a_\pm(t) = (a_1\pm a_2)/\sqrt{2}$. This result differs from Eq. \eqref{eq:app:solres} in that both $a_+$ and $a_-$ grow with the same rate; note above that this result holds only for $t\gg \frac{1}{d(0)} \log[2 d(0)/\varepsilon]$, so taking the limit $\varepsilon\to 0$ is non-trivial. Nonetheless, in that limit we recover the correct growth rate $a(t) \propto \exp\left(-\frac{\eta t}{2} + d\frac{1-e^{-\lambda t}}{\lambda} \right) $, suggesting that corrections are of the order 1.

We can now find the maximal value of $a$ (at least the exponential part) for a given $d$: $A(d,\lambda,\eta,\varepsilon) = {\rm max}_t e^{-\eta t/2 } e^{\int_0^t dt' \sqrt{d^2(t')-\varepsilon^2}}$ 
\begin{equation}
\begin{gathered}
        t_{max}= \frac{1}{2 \lambda} \log \frac{d^2}{\varepsilon^2+\eta^2/4},
        \\
        \sqrt{d^2(t_{max}) -\varepsilon^2} = \eta/2,
\end{gathered}
\end{equation}
where the second line implies the necessity of $\lambda\ll \eta$ for this solution to hold. The expression for $A(d,\lambda,\eta,\varepsilon)$ is much more cumbersome, but can be simplified with a Taylor expansion for $\varepsilon$ for $\varepsilon \ll \eta$, resulting in:
\begin{equation}
    A(d,\lambda,\eta,\varepsilon) \approx  A(d,\lambda,\eta,0)  e^{-\varepsilon^2\left(\frac{1}{\lambda \eta}- \frac{1}{2d \lambda}\right)},
\end{equation}
from which an estimate of FWHM can be obtained:
\begin{equation}
        \Delta \varepsilon^{FWHM}_{\lambda \ll \eta} 
    =
    2\sqrt{\frac{ \lambda \log 2}{\frac{1}{\eta}- \frac{1}{2d}}}.
\end{equation}

\bibliography{main.bbl} 

\end{document}